\documentclass{article}\usepackage{spconf,amsmath,graphicx}     % ICASSP
\ninept
\usepackage{graphics} % for pdf, bitmapped graphics files
\usepackage{amssymb}  % assumes amsmath package installed
\usepackage{amsmath}
\def\norm #1{\left\|#1\right\|}

\def\twon #1{\left\|#1\right\|_2}

\def\frobn #1{\left\|#1\right\|_{\text{F}}}

\def\atomn #1{\left\|#1\right\|_{\cA}}

\def\abs #1{\left|#1\right|}

\def\st{\text{subject to }}

\def\bC{\mathbb{C}}

\def\bT{\mathbb{T}}

\def\m #1{\boldsymbol{#1}}

\def\cA{\mathcal{A}}

\def\cD{\mathcal{D}}

\def\cM{\mathcal{M}}

\def\bee{\begin{equation}}
\def\ene{\end{equation}}

\def\beq{\begin{eqnarray}}
\def\enq{\end{eqnarray}}

\newtheorem{thm}{Theorem}

\def\equ #1{\begin{equation}#1\end{equation}}

\def\sbra #1{\left(#1\right)}
\def\mbra #1{\left[#1\right]}
\def\lbra #1{\left\{#1\right\}}

\def\tr #1{\text{tr}#1}

\def\rank #1{\text{rank}#1}
\def\st {\text{ subject to }}

\title{Achieving High Resolution for Super-resolution via Reweighted Atomic Norm Minimization}
\name{Zai Yang and Lihua Xie, Fellow, IEEE \thanks{Submitted to ICASSP 2015, Brisbane, Australia, April 2015.}}
\address{School of Electrical and Electronic Engineering\\
Nanyang Technological University, 639798, Singapore}

\begin{document}
\maketitle

%%%%%%%%%%%%%%%%%%%%%%%%%%%%%%%%%%%%%%%%%%%%%%%%%%%%%%%%%%%%%%%%%%%%%%%%%%%%%%%%
\begin{abstract}
The super-resolution theory developed recently by Cand\`{e}s and Fernandes-Granda aims to recover fine details of a sparse frequency spectrum from coarse scale information only. The theory was then extended to the cases with compressive samples and/or multiple measurement vectors. However, the existing atomic norm (or total variation norm) techniques succeed only if the frequencies are sufficiently separated, prohibiting commonly known high resolution. In this paper, a reweighted atomic-norm minimization (RAM) approach is proposed which iteratively carries out atomic norm minimization (ANM) with a sound reweighting strategy that enhances sparsity and resolution. It is demonstrated analytically and via numerical simulations that the proposed method achieves high resolution with application to DOA estimation.
\end{abstract}

%a novel nonconvex optimization method is proposed which guarantees exact recovery under no resolution limit. A locally convergent iterative algorithm is implemented to solve the nonconvex problem. The algorithm iteratively carries out ANM with a sound reweighting strategy which , and is termed as reweighted atomic-norm minimization (RAM). Extensive numerical simulations are carried out to demonstrate that the proposed method achieves high resolution.

%\begin{keywords}
%Continuous compressed sensing, high resolution, reweighted atomic norm minimization, super-resolution.
%\end{keywords}

\section{Introduction}
Frequency analysis of signals \cite{stoica2005spectral} is a classical problem that has broad applications ranging from communications, radar, array processing to seismology and astronomy. Grid-based sparse methods have been vastly studied in the past decade with the development of compressed sensing (CS) which exploit signal sparsity--the number of frequency components $K$ is small--but suffer from basis mismatches due to the need of gridding of the frequency interval \cite{chi2011sensitivity,yang2013off}. Its research has been recently advanced owing to the mathematical theory of super-resolution introduced by Cand\`{e}s and Fernandes-Granda \cite{candes2013towards}, which refers to recovery of fine details of a sparse frequency spectrum from coarse scale time-domain samples only. They propose a gridless atomic norm (or total variation norm) technique, which can be cast as semidefinite programming (SDP), and prove that a continuous frequency spectrum can be recovered with infinite precision given a set of $N$ regularly spaced samples. The technical method and theoretical result were then extended by Tang {\em et al.} \cite{tang2012compressed} to the case of partial/compressive samples, showing that only a number of $M=O\sbra{K\ln K\ln N}$ random samples are sufficient for the recovery with high probability via atomic norm minimization (ANM). Moreover, Yang and Xie \cite{yang2014continuous,yang2014exact} study the multiple-measurement-vector (MMV) case, which arises naturally in array processing applications, with similar results proven using extended MMV atomic norm methods. However, a major problem of existing atomic norm methods is that the frequency spectrum can be recovered only when the frequencies are sufficiently separated, prohibiting commonly known high resolution--the capability of resolving two closely spaced frequency components. A sufficient minimum separation of frequencies is $\frac{4}{N}$ in theory. Empirical evidences in \cite{tang2012compressed} suggest that this number can be reduced to $\frac{1}{N}$, while according to \cite{yang2014exact,yang2014gridless} it also depends on $K$, $M$ and the number of measurement vectors.

In this paper, we attempt to propose a high resolution gridless sparse method for super-resolution to break the resolution limit of existing atomic norm methods. Our method is motivated by the formulations and properties of atomic $\ell_0$ norm and the atomic norm in \cite{yang2014continuous,yang2014exact}. In particular, the atomic $\ell_0$ norm has no resolution limit but is NP hard to compute. To the contrary, as a convex relaxation the atomic norm can be efficiently computed but suffers from a resolution limit as mentioned above. We propose a novel sparse metric and theoretically show that the new metric fills the gap between the atomic $\ell_0$ norm and the atomic norm. It approaches the former under appropriate parameter setting. With the sparse metric we formulate a nonconvex optimization problem and present a locally convergent iterative algorithm for super-resolution. The algorithm iteratively carries out ANM with a sound reweighting strategy, which determines preference of frequency selection based on the latest estimate and enhances sparsity and resolution, and is termed as reweighted atomic-norm minimization (RAM). To the best of our knowledge, RAM implements the first reweighting strategy in the continuous dictionary setting while existing reweighted $\ell_1$ algorithms (see, e.g., \cite{candes2008enhancing}) are for the discrete setting. Extensive numerical simulations are carried out to demonstrate the high resolution performance of RAM with application to DOA estimation compared to existing arts.

\section{Preliminary Results} \label{sec:preliminary}

\subsection{Problem Formulation}
We consider the super-resolution problem in the most general case with partial samples and MMVs. In particular, we observe the samples of the data matrix $\m{Y}^o\in\bC^{N\times L}$ on the rows indexed by $\m{\Omega} \subset \mbra{N}\triangleq\lbra{1,2,\dots,N}$ of size $M=\abs{\m{\Omega}} < N$, denoted by $\m{Y}_{\m{\Omega}}^o$. The $(j,t)$th element of $\m{Y}^o$ is (corrupted by noise in practice)
\equ{y_{jt}^o=\sum_{k=1}^{K} \m{a}\sbra{f_k}\m{s}_k, \quad \sbra{j,t}\in \mbra{N}\times\mbra{L}, \label{formu:observmodel2}}
where $\m{a}\sbra{f}=\mbra{1,e^{i2\pi f},\dots,e^{i2\pi\sbra{N-1}f}}^T\in\bC^N$ denotes a discrete complex sinusoid with frequency $f\in\bT\triangleq\left[0,1\right]$, and $\m{s}_k\in\bC^{1\times L}$ is the coefficient vector of the $k$th sinusoid.
That is, each column of $\m{Y}^o$ is superimposed by $K$ discrete sinusoids. We are interested in recovering the frequencies $\lbra{f_k}$ given $\m{Y}_{\m{\Omega}}^o$. Meanwhile, it is also of interest to recover the complete data matrix $\m{Y}^o$. The resulting problem is known as continuous/off-grid CS according to \cite{tang2012compressed,yang2014continuous}, which differs from existing CS framework in the sense that every frequency $f_k$ can take any continuous value in $\bT$ rather than constrained on a finite discrete grid. The single-measurement-vector (SMV) case where $L=1$ is known as line spectral estimation. The MMV case where $L>1$ is common in array processing. Therein the sampling index set $\m{\Omega}$ refers to sensor placement of a linear sensor array and a smaller sample size means use of less sensors. $\m{Y}_{\m{\Omega}}^o$ consists of measurements of the sensor array and each column vector corresponds to one data snapshot. Each frequency corresponds to the direction of one source. Therefore, the frequency estimation problem is known as direction of arrival (DOA) estimation.

\subsection{Existing Gridless Sparse Methods}

The super-resolution or continuous CS problem is tackled from the perspective of signal recovery. The frequencies are then retrieved from the computational result. In particular, we seek a \emph{frequency-sparse} candidate $\m{Y}$, which is composed of a few frequency components, in a feasible domain defined by the observed samples. To do this, we first define a sparse metric of $\m{Y}$ and then optimize the metric over the feasible domain. A direct sparse metric is the smallest number of frequency components composing $\m{Y}$, known as the atomic $\ell_0$ norm and denoted by $\norm{\m{Y}}_{\cA,0}$. According to \cite{tang2012compressed,yang2014continuous,yang2014exact} $\norm{\m{Y}}_{\cA,0}$ can be characterized as the following rank minimization problem:
\equ{\begin{split}\norm{\m{Y}}_{\cA,0}
=&\min_{\m{u}} \rank\sbra{T\sbra{\m{u}}},\\
&\st \tr\sbra{\m{Y}^HT\sbra{\m{u}}^{-1}\m{Y}}<+\infty,\\
&\phantom{\st } T\sbra{\m{u}}\geq\m{0}.\end{split} \label{formu:atom0norm}}
The first constraint in (\ref{formu:atom0norm}) imposes that $\m{Y}$ lies in the range space of a (Hermitian) Toeplitz matrix $T\sbra{\m{u}}\in\bC^{N\times N}$ whose first row is specified by the transpose of $\m{u}\in\bC^{N}$.
The frequencies composing $\m{Y}$ are encoded in $T\sbra{\m{u}}$. Once an optimizer of $\m{u}$, say $\m{u}^*$, is obtained the frequencies can be retrieved from $T\sbra{\m{u}^*}$ according to the Vandermonde decomposition lemma (see, e.g., \cite{stoica2005spectral}), which states that any positive semidefinite (PSD) Toeplitz matrix $T\sbra{\m{u}^*}$ can be decomposed as $T\sbra{\m{u}^*}=\sum_{k=1}^{K^*} p_k^*\m{a}\sbra{f_k^*}\m{a}\sbra{f_k^*}^H$,
where the order $K^*=\rank\sbra{T\sbra{\m{u}^*}}$ and $p_k^*>0$ (see a method for realization of the decomposition in \cite[Appendix A]{yang2014gridless}). The atomic $\ell_0$ norm directly enhances sparsity, however, it is nonconvex and NP-hard to compute and encourages computationally feasible alternatives. In this spirit, the atomic ($\ell_1$) norm, denoted by $\atomn{\m{Y}}$, is introduced as a convex relaxation of $\norm{\m{Y}}_{\cA,0}$ and has the following semidefinite formulation \cite{tang2012compressed,yang2014continuous,yang2014exact}:
\equ{\begin{split}\norm{\m{Y}}_{\cA}=
&\min_{\m{u}} \frac{1}{2\sqrt{N}} \mbra{\tr\sbra{T\sbra{\m{u}}} + \tr\sbra{\m{Y}^HT\sbra{\m{u}}^{-1} \m{Y}}},\\
&\st T\sbra{\m{u}}\geq\m{0}. \end{split} \label{formu:AN_SDP}}
From the perspective of low rank matrix recovery (LRMR), (\ref{formu:AN_SDP}) attempts to recover the low rank matrix $T\sbra{\m{u}}$ by relaxing the pseudo-rank norm in (\ref{formu:atom0norm}) to the nuclear norm or equivalently the trace norm for a PSD matrix. The atomic norm is advantageous in computation compared to the atomic $\ell_0$ norm, however, it suffers from a resolution limit due to the relaxation which is not shared by the latter \cite{candes2013towards,tang2012compressed,yang2014exact}.

\section{Enhancing Sparsity and Resolution via A Novel Sparse Metric} \label{sec:novelmetric}

\begin{figure}
\centering
  \includegraphics[width=2.9in]{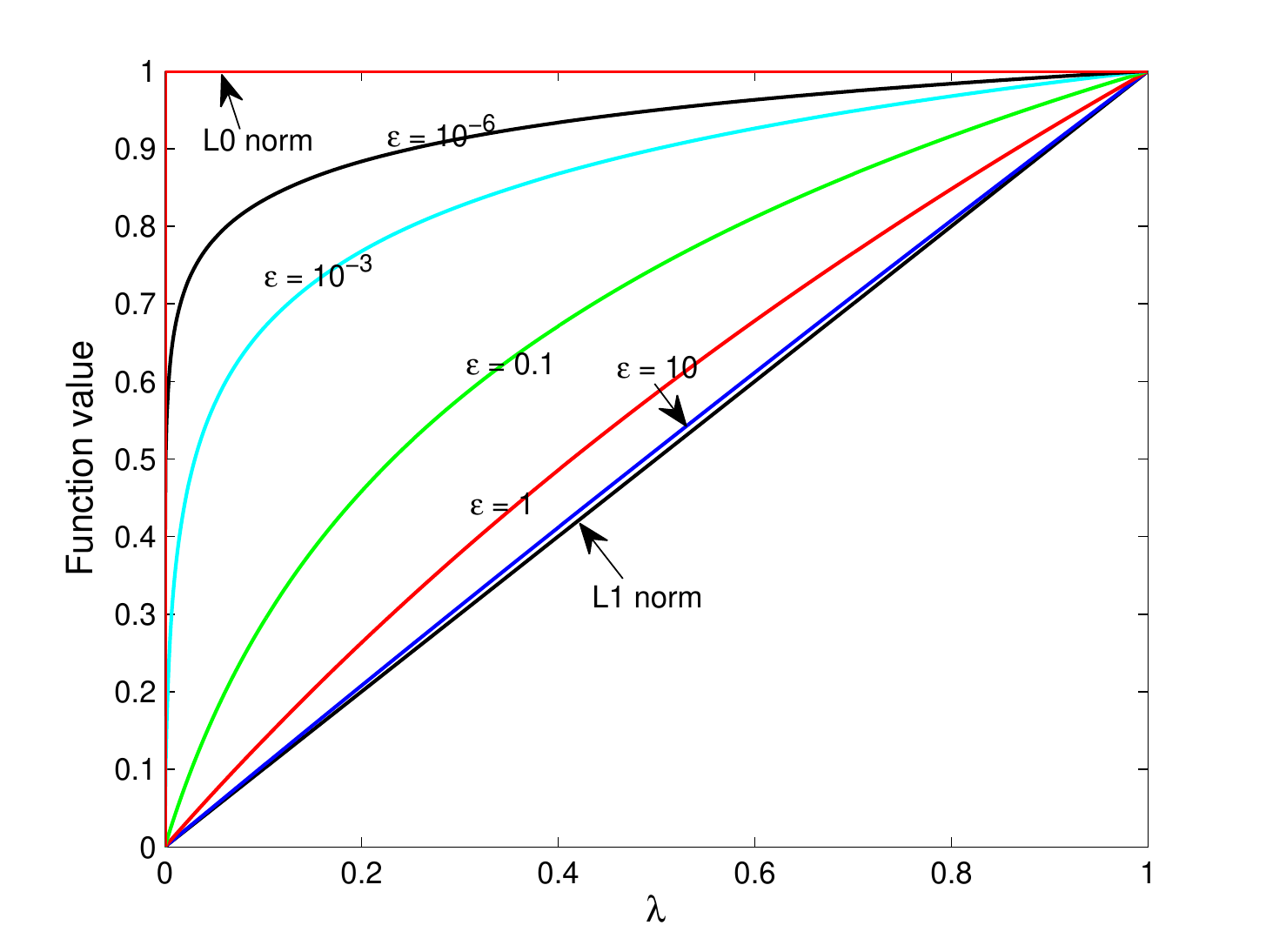}
\centering
\caption{The sparsity-promoting property of $\cM^{\epsilon}\sbra{\cdot}$ with different $\epsilon$. The plotted curves include the $\ell_0$ and $\ell_1$ norms corresponding to $\norm{\cdot}_{\cA,0}$ and $\norm{\cdot}_{\cA}$ respectively, and $\ln\abs{\lambda+\epsilon}$ corresponding to $\cM^{\epsilon}\sbra{\cdot}$ with $\epsilon=10, 1, 0.1, 10^{-3}$ and $10^{-6}$. $\ln\abs{\lambda+\epsilon}$ is translated and scaled such that it equals 0 and 1 at $\lambda=0$ and 1 respectively for better illustration.} \label{Fig:sparsity}
\end{figure}

Inspired by the link between continuous CS and LRMR demonstrated above, we propose the following sparse metric of $\m{Y}$:
\equ{\begin{split}\cM^{\epsilon}\sbra{\m{Y}}=
&\min_{\m{u}} \ln\abs{T\sbra{\m{u}}+\epsilon\m{I}} + \tr\sbra{\m{Y}^HT\sbra{\m{u}}^{-1}\m{Y}},\\
&\st T\sbra{\m{u}}\geq\m{0},\end{split} \label{formu:nonconvexrelax}}
where $\epsilon>0$ is a regularization parameter that avoids the first term being $-\infty$ when $T\sbra{\m{u}}$ is rank deficient.
Note that the log-det heuristic $\ln\abs{\cdot}$ is a common smooth surrogate for the matrix rank (see, e.g., \cite{fazel2003log}). From the perspective of LRMR, the atomic $\ell_0$ norm minimizes the number of nonzero eigenvalues of $T\sbra{\m{u}}$ while the atomic norm minimizes the sum of the eigenvalues. In contrast, the new metric $\cM^{\epsilon}\sbra{\m{Y}}$ penalizes $\sum_{k=1}^N\ln\abs{\lambda_k+\epsilon}$, where $\lbra{\lambda_k}_{k=1}^N$ denotes the eigenvalues. We plot the function $h(\lambda)=\ln\abs{\lambda+\epsilon}$ with different $\epsilon$'s in Fig. \ref{Fig:sparsity}, according to which we expect that the new metric $\cM^{\epsilon}\sbra{\m{Y}}$ bridges $\atomn{\m{Y}}$ and $\norm{\m{Y}}_{\cA,0}$ when $\epsilon$ varies from $+\infty$ to $0$. Formally, we have the following results and we provide their proofs in an extended journal paper \cite{yang2014enhancing}.

\begin{thm} Let $\epsilon\rightarrow+\infty$. Then,
\equ{\cM^{\epsilon}\sbra{\m{Y}}-N\ln\epsilon \sim 2\sqrt{N}\atomn{\m{Y}}\epsilon^{-\frac{1}{2}}, \label{formu:epsilontoinf}}
i.e., they are equivalent infinitesimals. \label{thm:epsilontoinf}
\end{thm}

\begin{thm} Let $\epsilon\rightarrow0$. Then, we have the following results:
\begin{enumerate}
 \item If $\norm{\m{Y}}_{\cA,0}\leq N-1$, then
     \equ{\cM^{\epsilon}\sbra{\m{Y}}\sim \sbra{\norm{\m{Y}}_{\cA,0}-N}\ln\frac{1}{\epsilon}, \label{formu:equivinf}}
     i.e., they are equivalent infinities. Otherwise, $\cM^{\epsilon}\sbra{\m{Y}}$ is a positive constant depending only on $\m{Y}$;
 \item Let $\m{u}_{\epsilon}^*$ be the optimizer of $\m{u}$ to the optimization problem in (\ref{formu:nonconvexrelax}). Then, the smallest $N-\norm{\m{Y}}_{\cA,0}$ eigenvalues of $T\sbra{\m{u}_{\epsilon}^*}$ are either zero or approach zero as fast as $\epsilon$;
 \item For any cluster point of $\m{u}_{\epsilon}^*$ at $\epsilon=0$, denoted by $\m{u}_0^*$, there exists an atomic decomposition $\m{Y}=\sum_{k=1}^K\m{a}\sbra{f_k}\m{s}_k$ of order $K=\norm{\m{Y}}_{\cA,0}$ such that $T\sbra{\m{u}_0^*}=\sum_{k=1}^K\twon{\m{s}_k}^2\m{a}\sbra{f_k}\m{a}\sbra{f_k}^H$.
\end{enumerate} \label{thm:epsilontozero}
\end{thm}

Theorem \ref{thm:epsilontoinf} shows that the new metric $\cM^{\epsilon}\sbra{\m{Y}}$ plays the same role as $\atomn{\m{Y}}$ in the limiting scenario when $\epsilon\rightarrow+\infty$, while Theorem \ref{thm:epsilontozero} says that it is equivalent to $\norm{\m{Y}}_{\cA,0}$ as $\epsilon\rightarrow0$. Consequently, it fills the gap between $\atomn{\m{Y}}$ and $\norm{\m{Y}}_{\cA,0}$ and enhances sparsity and resolution compared to $\atomn{\m{Y}}$ as $\epsilon$ gets small. Moreover, Theorem \ref{thm:epsilontozero} characterizes the properties of the optimizer $\m{u}_{\epsilon}^*$ as $\epsilon\rightarrow0$ including the convergent speed of the smallest $N-K$ eigenvalues and the limiting form of $T\sbra{\m{u}^*}$ via the Vandermonde decomposition. In fact, we always observe via simulations that the smallest $N-K$ eigenvalues of $T\sbra{\m{u}^*}$ become zero once $\epsilon$ is modestly small.

\section{Reweighted Atomic-Norm Minimization} \label{sec:RAM}

\subsection{A Locally Convergent Iterative Algorithm}
With the proposed sparse metric $\cM^{\epsilon}\sbra{\m{Y}}$, we solve the following optimization problem for signal and frequency recovery:
\equ{\begin{split}
\min_{\m{Y}}\cM^{\epsilon}\sbra{\m{Y}}, \st \m{Y}\in\cD, \end{split} \label{formu:criterion3}}
or equivalently,
\equ{\begin{split}
&\min_{\m{Y},\m{u}} \ln\abs{T\sbra{\m{u}}+\epsilon\m{I}} + \tr\sbra{\m{Y}^HT\sbra{\m{u}}^{-1}\m{Y}},\\
&\st T\sbra{\m{u}}\geq\m{0} \text{ and } \m{Y}\in\cD, \end{split} \label{formu:problem}}
where $\cD$ denotes the feasible domain of $\m{Y}$. For example, in the noiseless case, it is the set $\lbra{\m{Y}: \m{Y}_{\m{\Omega}}=\m{Y}_{\m{\Omega}}^o}$.
Since the log-det term $\ln\abs{T\sbra{\m{u}}+\epsilon\m{I}}$ is a concave function of $\m{u}$, the problem is nonconvex and no efficient algorithms can guarantee to obtain the global optimum. A majorization-maximization (MM) algorithm is introduced as follows. Let $\m{u}_j$ denote the $j$th iterate of the optimization variable $\m{u}$. Then, at the $\sbra{j+1}$th iteration we replace $\ln\abs{T\sbra{\m{u}}+\epsilon\m{I}}$ by its tangent plane at the current value $\m{u}=\m{u}_j$. As a result, the optimization problem at the $\sbra{j+1}$th iteration becomes
\equ{\begin{split}
&\min_{\m{Y},\m{u}} \tr\mbra{\sbra{T\sbra{\m{u}_j}+\epsilon\m{I}}^{-1}T\sbra{\m{u}}} + \tr\sbra{\m{Y}^HT\sbra{\m{u}}^{-1}\m{Y}},\\
&\st T\sbra{\m{u}}\geq\m{0} \text{ and } \st \m{Y}\in\cD. \end{split} \label{formu:problem_j}}
Since $\ln\abs{T\sbra{\m{u}}+\epsilon\m{I}}$ is strictly concave in $\m{u}$, at each iteration its value decreases by an amount greater than the decrease of its tangent plane. It follows that the objective function in (\ref{formu:problem}) monotonically decreases at each iteration and converges to a local minimum.

\subsection{Interpretation as RAM }\label{sec:WAN_SDP}
To interpret the optimization problem in (\ref{formu:problem_j}), let us define a \emph{weighted continuous dictionary}
\equ{\cA^w\triangleq\lbra{\m{a}^w\sbra{f}= w\sbra{f}\m{a}\sbra{f}:\; f\in\bT}}
w.r.t. the original continuous dictionary $\lbra{\m{a}\sbra{f}:\; f\in\bT}$, where $w\sbra{f}\geq 0$ is a weighting function. For $\m{Y}\in\bC^{N\times L}$, we define its \emph{weighted atomic norm} w.r.t. $\cA^w$ as its atomic norm induced by $\cA^w$:
\equ{\begin{split}
&\norm{\m{Y}}_{\cA^w}\triangleq\inf\lbra{\sum_k \twon{\m{s}^w_k}: \m{Y} = \sum_k \m{a}^w\sbra{f_k}\m{s}^w_k, f_k\in\bT}\\
&=\inf\lbra{\sum_k w\sbra{f_k}^{-1}\twon{\m{s}_k}: \m{Y} = \sum_k \m{a}\sbra{f_k}\m{s}_k, f_k\in\bT }. \end{split}}
According to the definition above, $w\sbra{f}$ specifies preference of the atoms $\lbra{\m{a}\sbra{f}}$. To be specific, an atom $\m{a}\sbra{f_0}$, $f_0\in\bT$, is more likely selected if $w\sbra{f_0}$ is larger. Moreover, the atomic norm is a special case of the weighted atomic norm with a constant weighting function (i.e., without any preference) according to \cite{yang2014continuous,yang2014exact}.

\begin{thm} Suppose that $w\sbra{f}= \frac{1}{\sqrt{\m{a}\sbra{f}^H\m{W}\m{a}\sbra{f}}}$ with $\m{W}\in\bC^{N\times N}$. Then,
\equ{\begin{split}
\norm{\m{Y}}_{\cA^w}=&\min_{\m{u}} \frac{\sqrt{N}}{2}\tr\sbra{\m{W}T\sbra{\m{u}}} + \frac{1}{2\sqrt{N}}\tr\sbra{\m{Y}^HT\sbra{\m{u}}^{-1}\m{Y}},\\
&\st T\sbra{\m{u}}\geq\m{0}. \end{split}} \label{thm:weightAN}
\end{thm}
%\begin{proof} See Appendix \ref{sec:Append_weightAN}.
%\end{proof}

Let $\m{W}_j=\frac{1}{N}\sbra{T\sbra{\m{u}_j}+\epsilon\m{I}}^{-1}$ and $w_j\sbra{f} = \frac{1}{\sqrt{\m{a}\sbra{f}^H \m{W}_j \m{a}\sbra{f}}}$. By Theorem \ref{thm:weightAN} we can rewrite the optimization problem in (\ref{formu:problem_j}) as the following \emph{weighted atomic norm minimization} problem:
\equ{\begin{split}
\min_{\m{Y}} \norm{\m{Y}}_{\cA^{w_j}},\st \st \m{Y}\in\cD. \end{split} \label{formu:WAN_j}}
As a result, the proposed iterative algorithm can be interpreted as \emph{reweighted atomic-norm minimization} (RAM). If we let $w_0(f)$ be a constant function or equivalently, $\m{u}_0=\m{0}$, such that there is no preference of the atoms at the first iteration, then the first iteration coincides with the ANM. From the second iteration, the preference is defined by the weighting function $w_j\sbra{f}$ specified above. Note that $w_j^2(f)$ corresponds to the power spectrum of Capon's beamforming (see, e.g., \cite{stoica2005spectral}) if $T\sbra{\m{u}_j}$ is interpreted as the covariance of the noiseless data and $\epsilon$ as the noise variance. Therefore, the reweighting strategy makes the frequencies around those estimated by the current iteration preferable at the next iteration and thus enhances sparsity. At the same time, the preference leads to finer details of the frequency spectrum in that area and enhances resolution. Since the ``noise variance'' $\epsilon$ can be translated as the confidence level in the current estimate, from this perspective we should gradually decrease $\epsilon$ and correspondingly increase the confidence in the solution during the algorithm.

\section{Numerical Simulations} \label{sec:simulation}

\subsection{Sparsity-Separation Phase Transition}
In this subsection, we study the success rate of RAM in super-resolution compared to ANM. In particular, we fix $N=64$ and $M=30$ with the sampling index set $\m{\Omega}$ being generated uniformly at random. We vary the duo $\sbra{K,\Delta_f}$ and at each combination we randomly generate $K$ frequencies such that they are mutually separated by at least $\Delta_f$. We randomly generate the amplitudes $\lbra{s_{kt}}$ independently and identically from a standard complex normal distribution. After obtaining the noiseless samples, we carry out super-resolution using ANM and RAM, both implemented by an off-the-shelf SDP solver SDPT3 \cite{toh1999sdpt3}. The recovery is called successful if both the relative MSE of signal recovery and the MSE of frequency recovery are less than $1\times 10^{-12}$. At each combination $\sbra{K,\Delta_f}$, the success rate is measured over 20 Monte Carlo runs. In RAM, we first scale the measurements such that $\frobn{\m{Y}_{\m{\Omega}}}^2=M$ and compensate the recovery afterwards. We start with $\m{u}_0=\m{0}$ and $\epsilon=1$ as default. We halve $\epsilon$ when beginning a new iteration until $\epsilon=\frac{1}{2^{10}}$. We terminate RAM if the relative change (in the Frobenius norm) of the solution $\m{Y}^*$ at two consecutive iterations is less than $1\times10^{-6}$ or the maximum number of iterations, set to 20, is reached.

\begin{figure}
\centering
\includegraphics[width=1.65in]{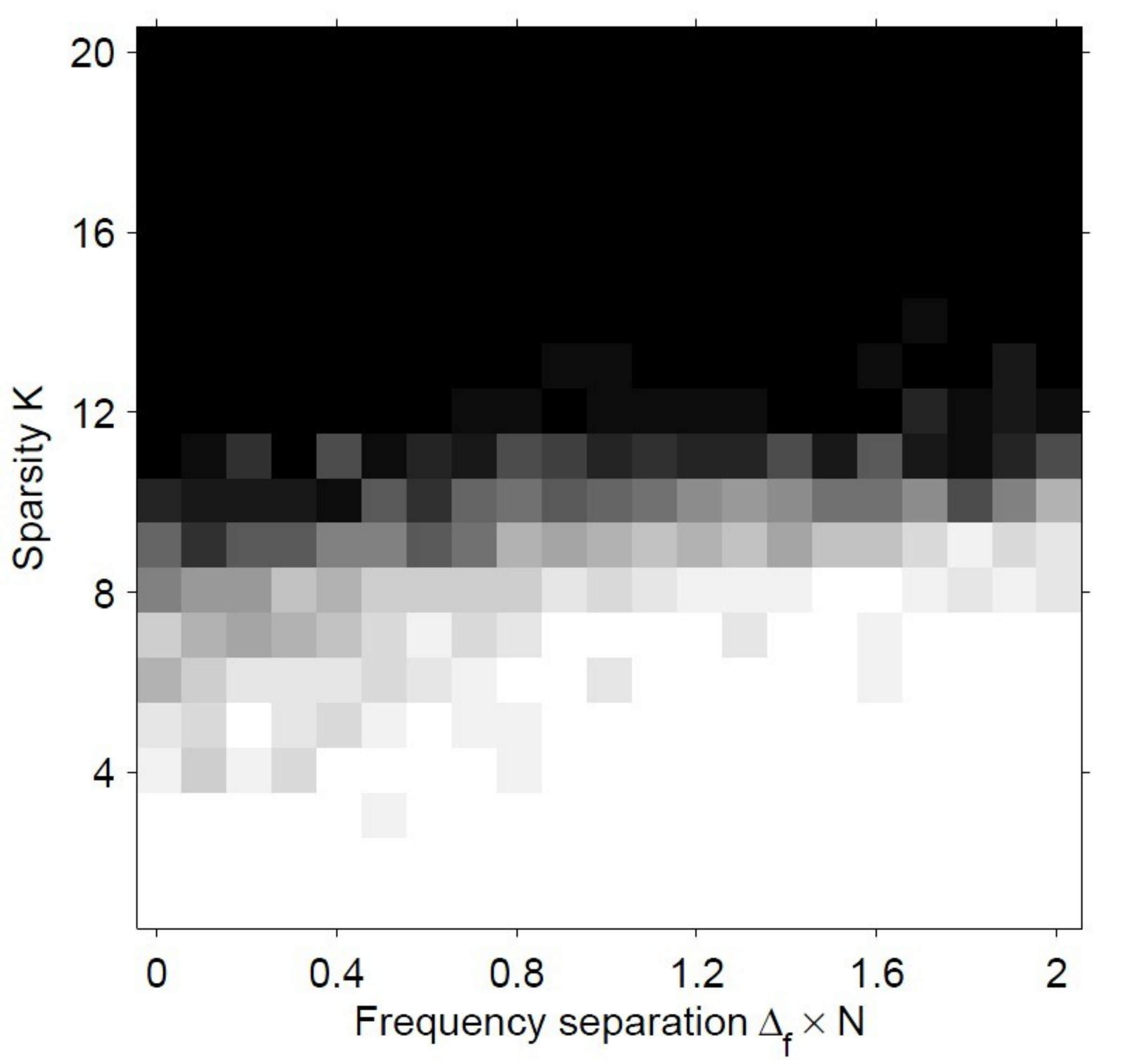} %
\includegraphics[width=1.63in]{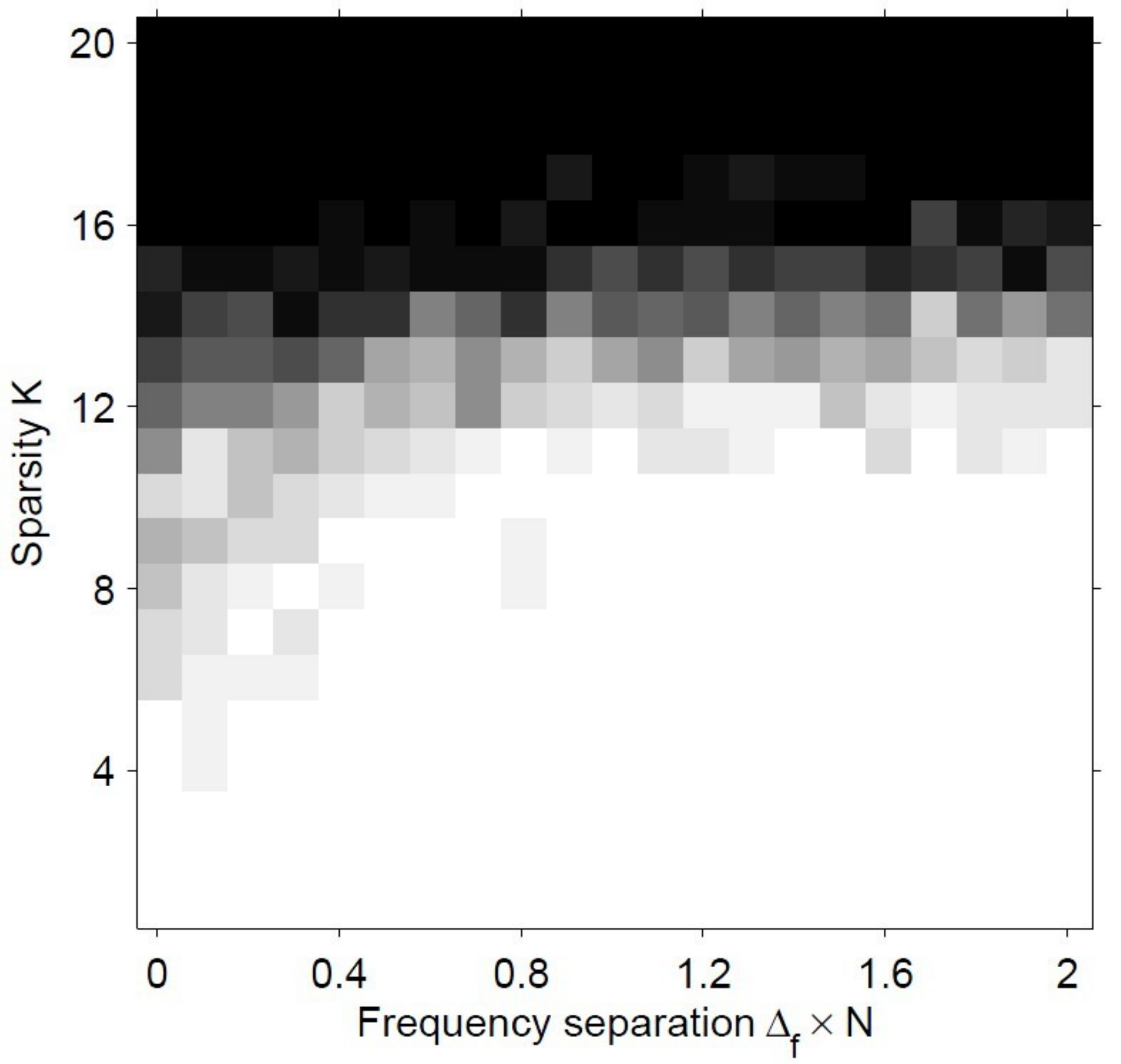}
\includegraphics[width=1.65in]{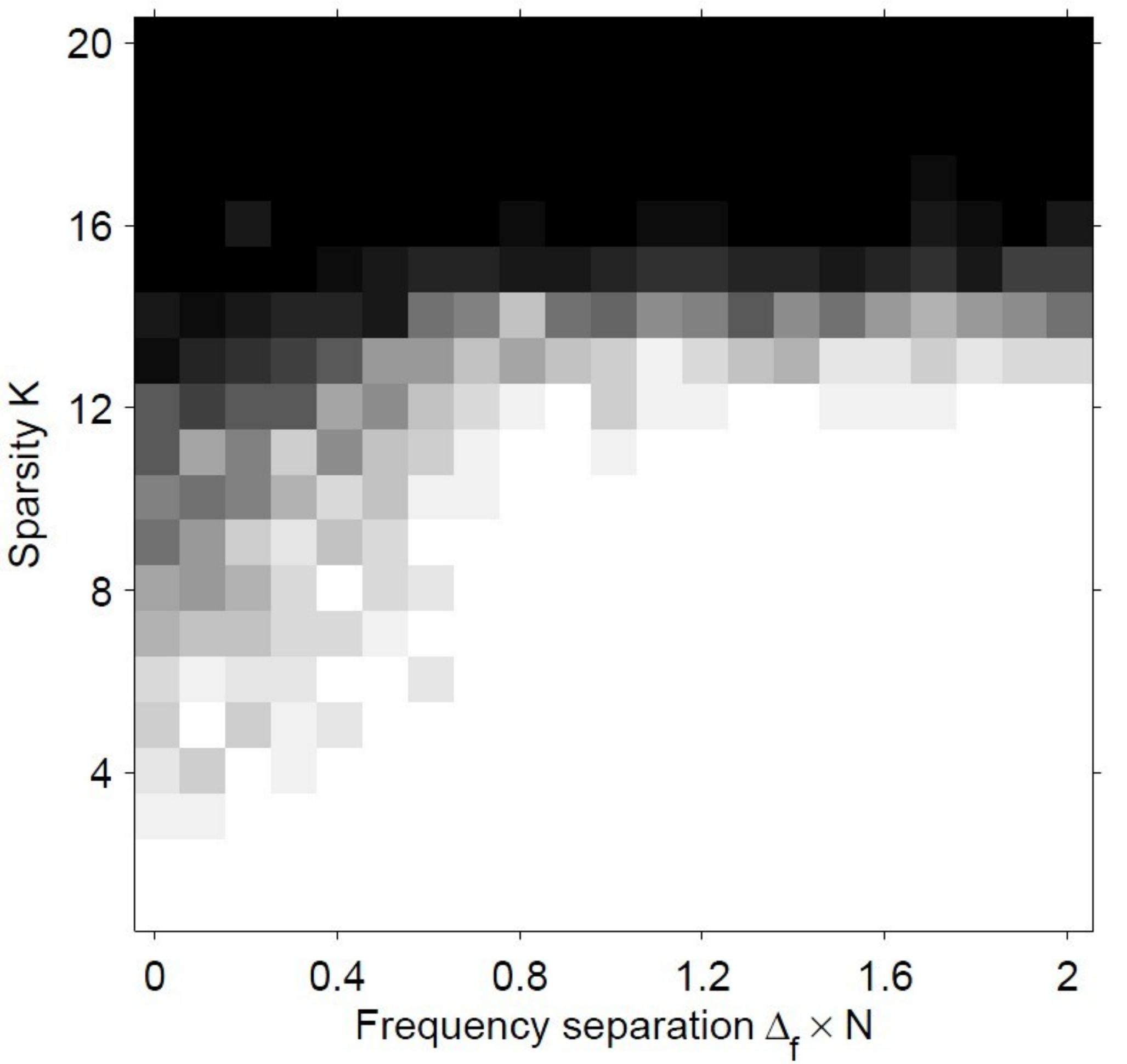} %
\includegraphics[width=1.63in]{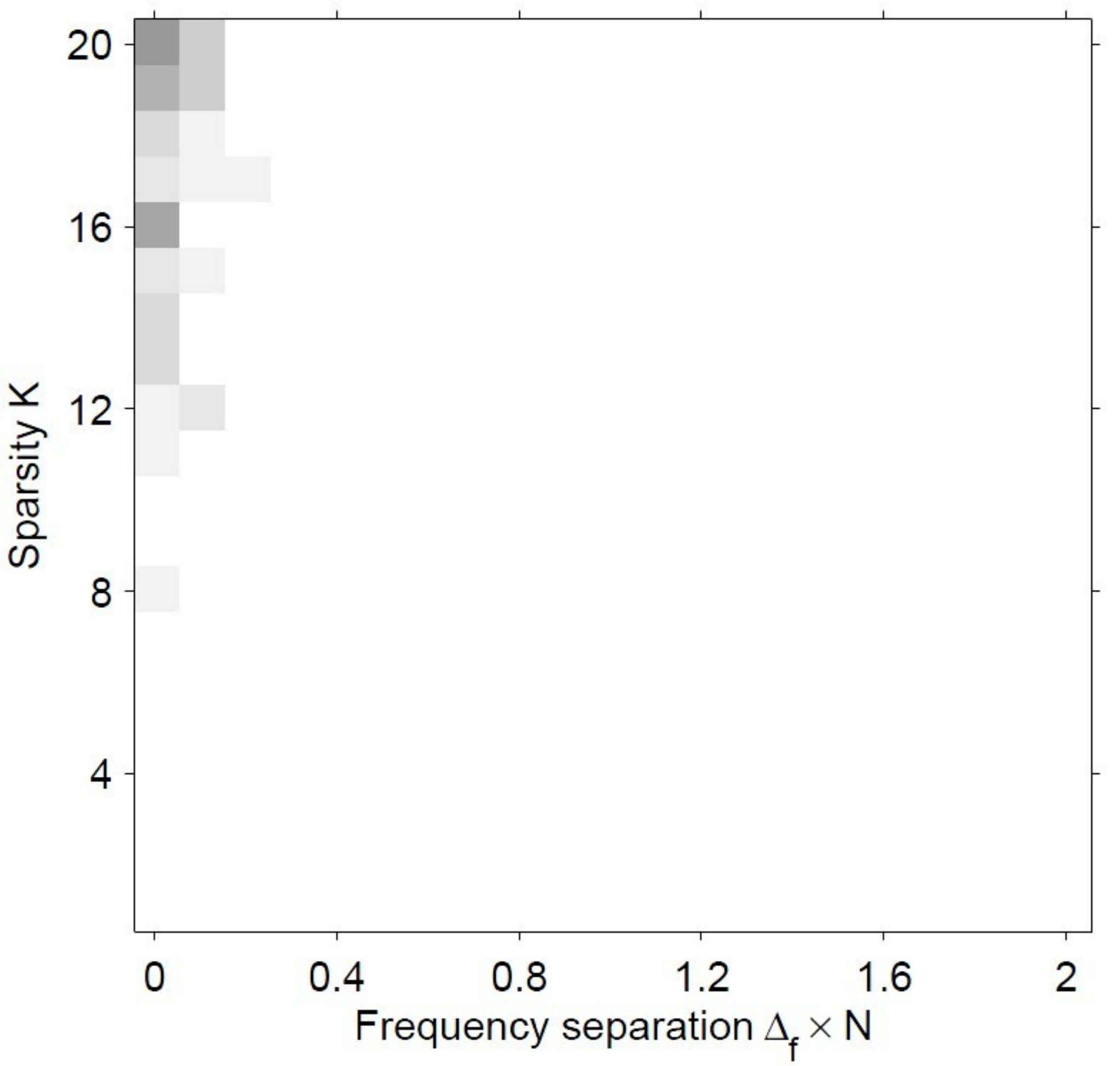}
\caption{Sparsity-separation phase transitions of ANM (left) and RAM (right) with $L=1$ (top) and $L=5$ (bottom), $N=64$ and $M=30$. The grayscale images present the success rates, where white and black colors indicate complete success and complete failure, respectively.} \label{Fig:phasetrans_1}
\end{figure}

We plot the success rates of ANM and RAM with $L=1,5$ in Fig. \ref{Fig:phasetrans_1}, where it is shown that successful recovery can be obtained with more ease with a smaller $K$ and a larger frequency separation $\Delta_f$, leading to a phase transition in the sparsity-separation domain. It is shown that RAM significantly enlarges the success phase and hence enhances sparsity and resolution compared to ANM. At $L=5$ we did not find a single failure in our simulation whenever $K\leq20$ and $\Delta_f\geq\frac{0.3}{N}$. The phase transitions of both ANM and RAM are not sharp since the frequencies are separated by \emph{at least} $\Delta_f$ and a set of \emph{well separated} frequencies can be possibly generated at a small value of $\Delta_f$. It is also observed that RAM tends to converge in less iterations with a smaller $K$ and a larger $\Delta_f$.

\subsection{Application to DOA Estimation} \label{sec:simulation_DOA}
We apply the proposed RAM method to DOA estimation. In particular, we consider a 10-element sparse linear array (SLA) with sensors' positions indexed by $\m{\Omega}=\lbra{1,2,5,6,8,12,15,17,19,20}$, where the distance between the first two sensors is half the wavelength. Hence, we have that $N=20$ and $M=10$. We consider that $K=4$ narrowband sources impinge on the sensor array from directions corresponding to frequencies $0.1$, $0.11$, $0.2$ and $0.5$, and powers $10$, $10$, $3$ and $1$, respectively. It is challenging to separate the first two sources which are separated by only $\frac{0.2}{N}$. Complex normal noise is added to the samples with variance $\sigma^2=1$ and $\cD$ is defined as $\lbra{\m{Y}: \frobn{\m{Y}_{\m{\Omega}}- \m{Y}_{\m{\Omega}}^o}\leq \eta^2}$, where $\eta^2=\sbra{ML+2\sqrt{ML}}\sigma^2$ (mean + twice standard deviation) upper bounds the noise energy with high probability. We consider both the cases of uncorrelated and correlated sources while the later case is usually considered to be more difficult with existing methods such as MUSIC (see, e.g., \cite{stoica2005spectral}). In the latter case, sources 1 and 3 are set to be coherent (completely correlated). Assume that $L=200$ data snapshots are collected which are corrupted by i.i.d. Gaussian noise of unit variance. We propose a dimension reduction technique to reduce the order of the SDP matrix from $L+N$ to $M+N$ and accelerate the computational speed, which is detailed in \cite{yang2014enhancing}. We terminate RAM within maximally 10 iterations and consider MUSIC and ANM for comparison.

\begin{figure}
\centering
\includegraphics[width=1.65in]{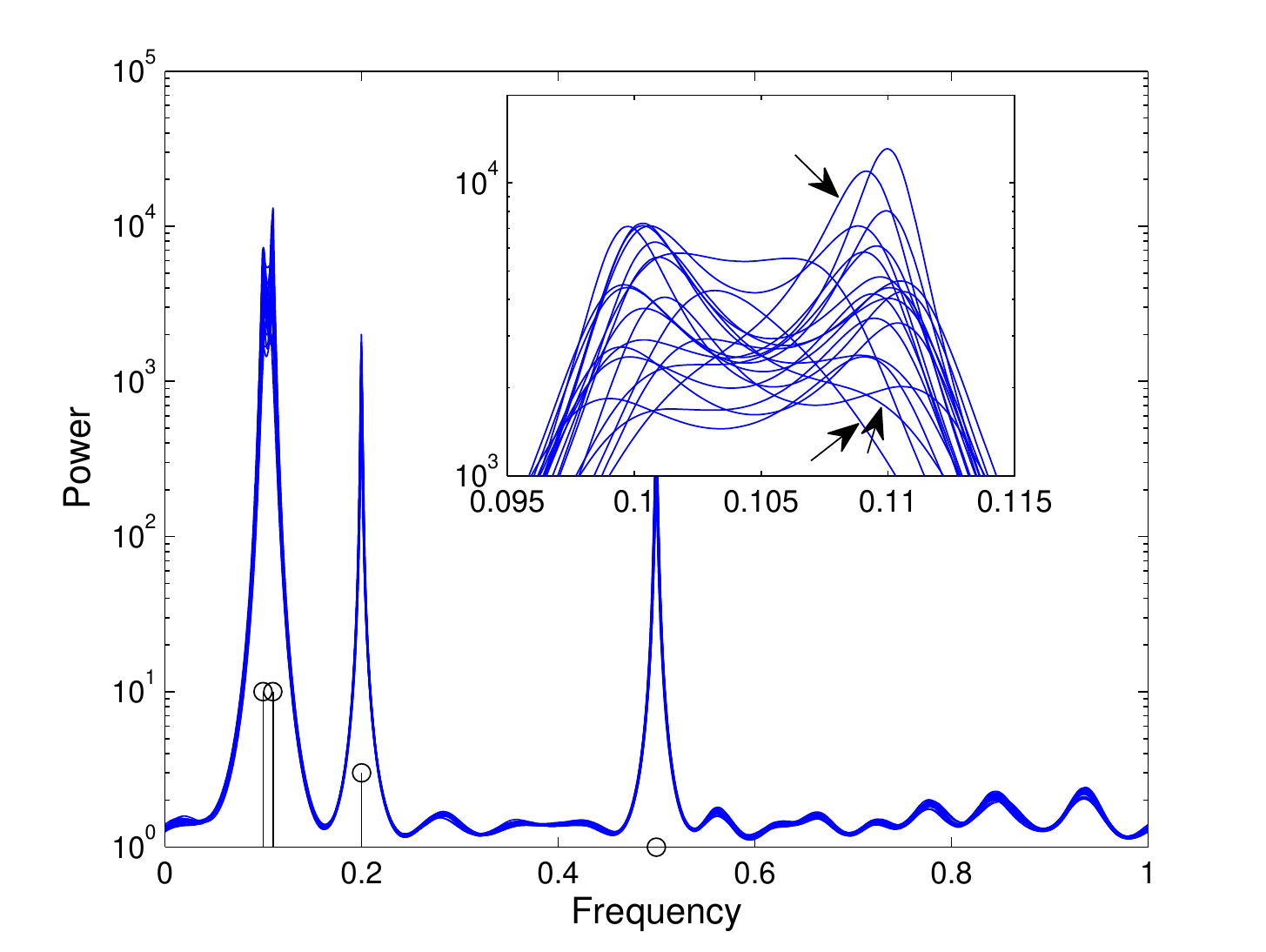} %
\includegraphics[width=1.63in]{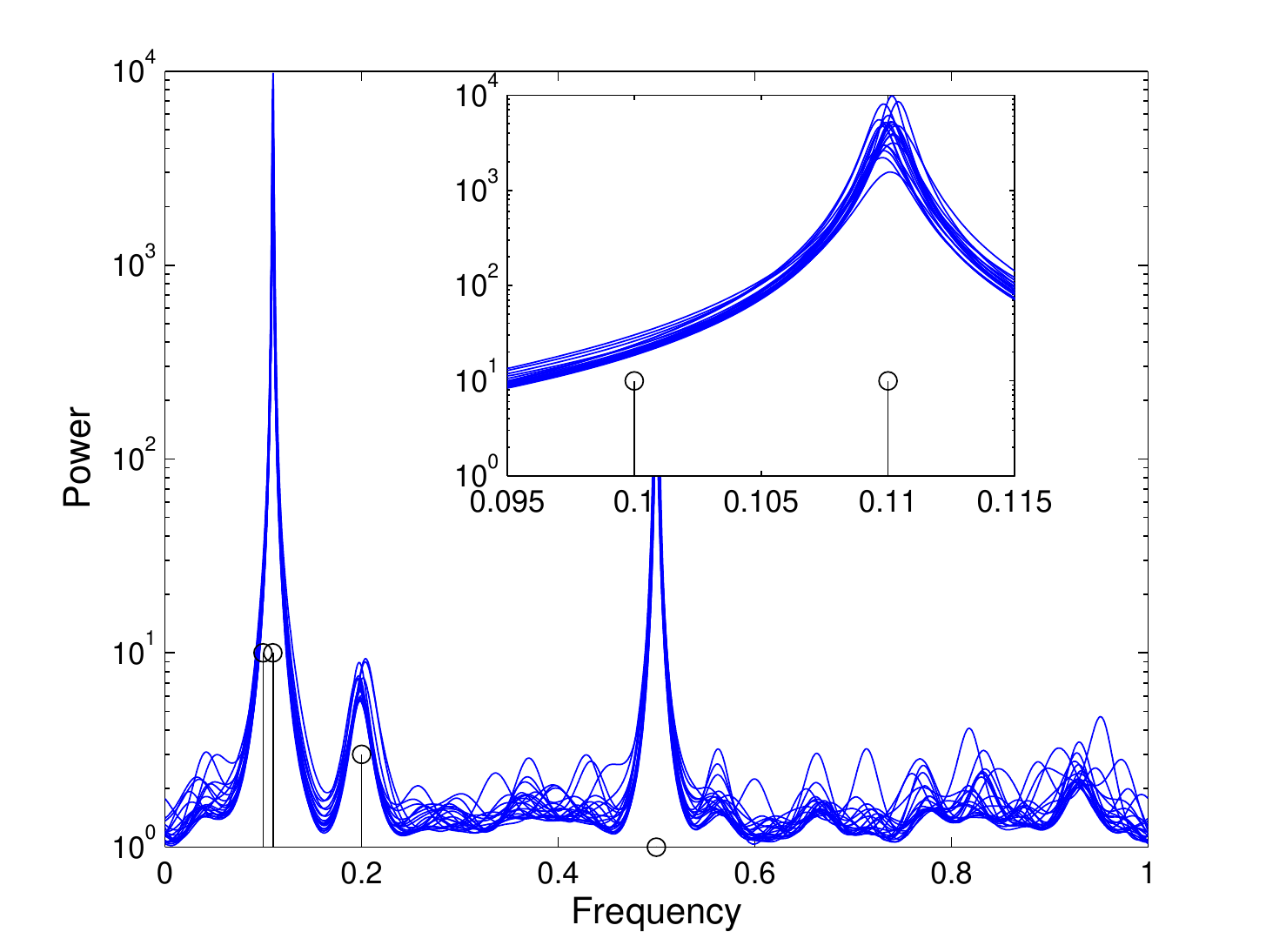}
\includegraphics[width=1.65in]{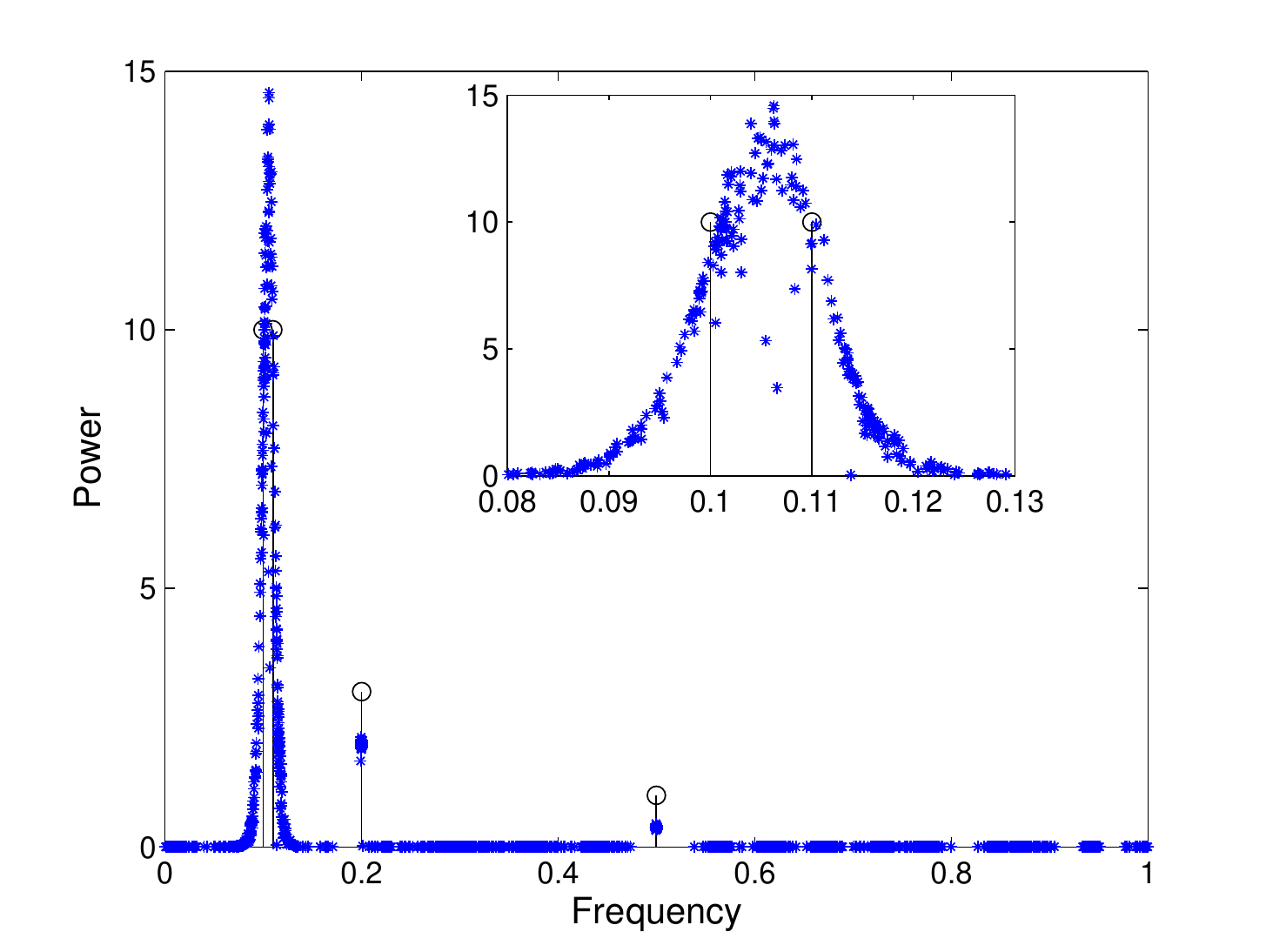} %
\includegraphics[width=1.63in]{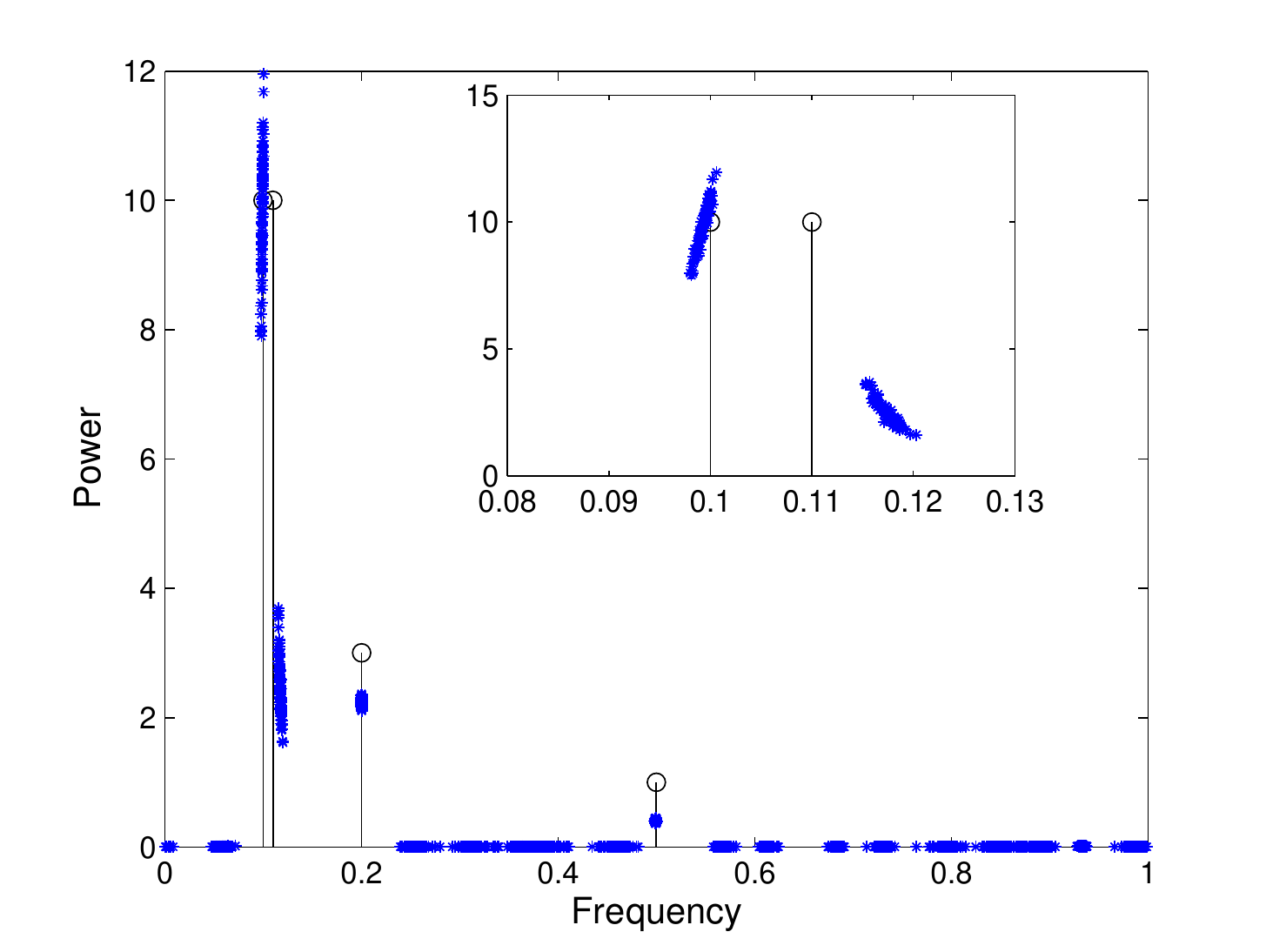}
\includegraphics[width=1.65in]{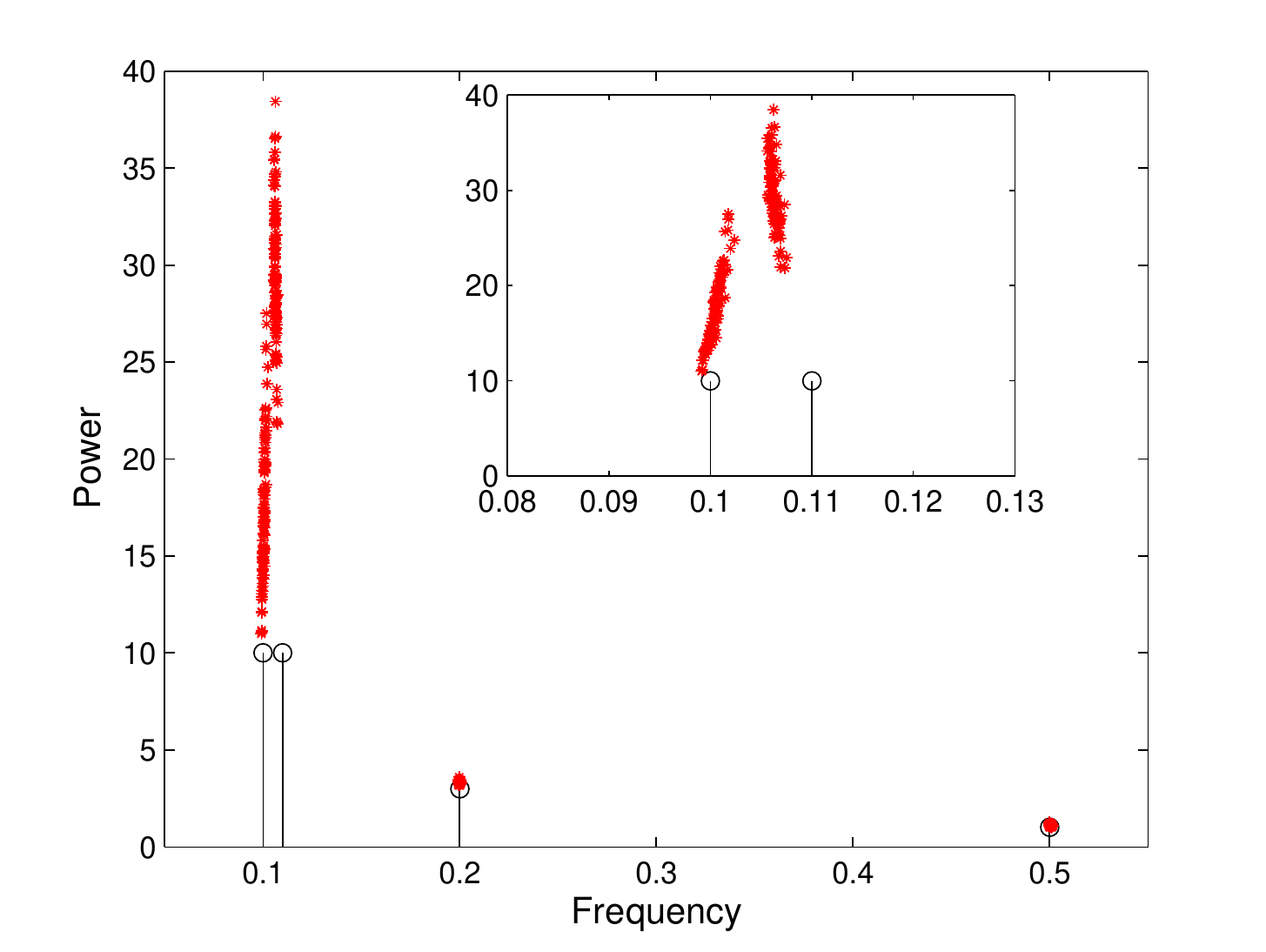} %
\includegraphics[width=1.63in]{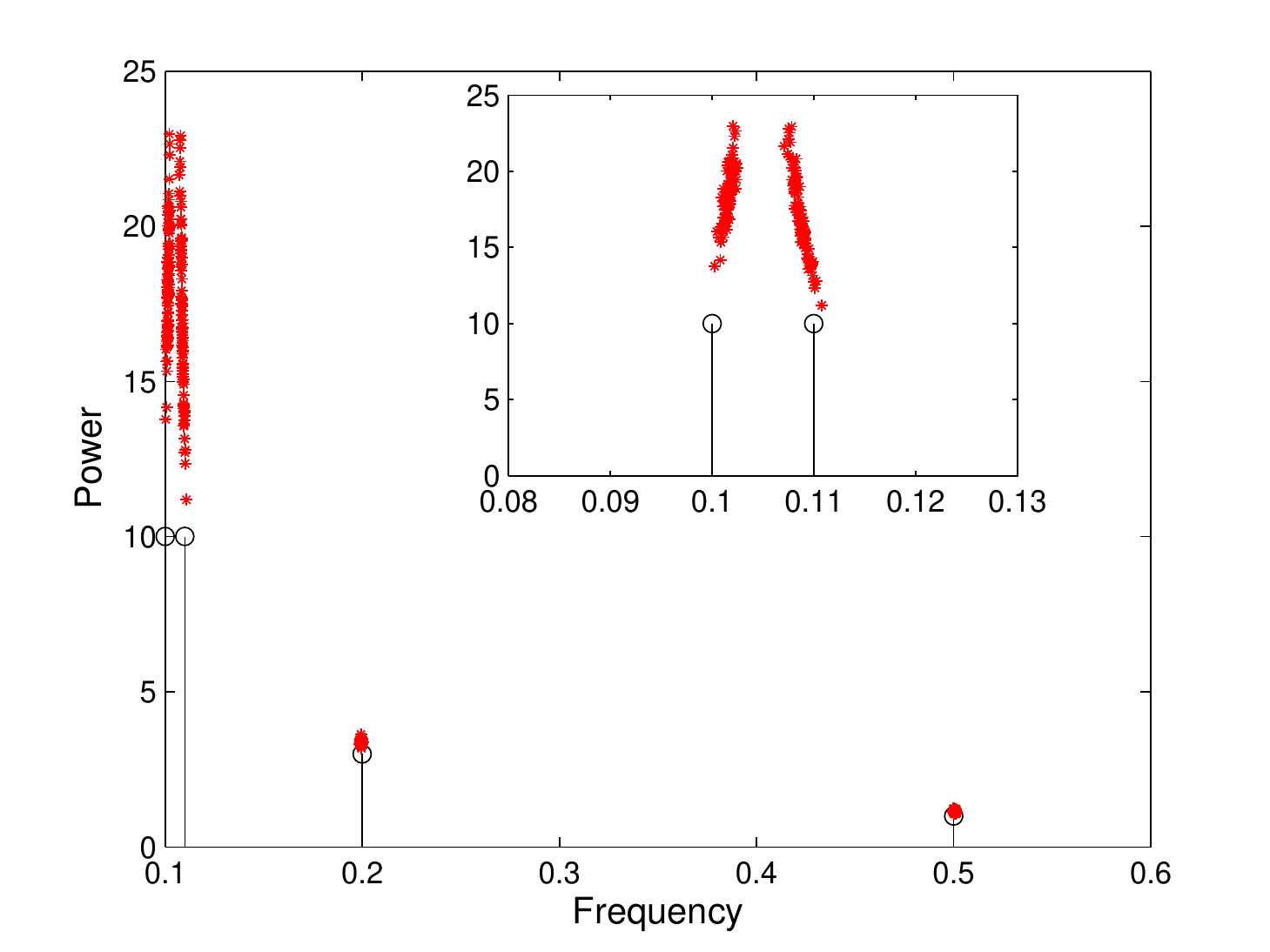}
\centering
\caption{Results of MUSIC (top), ANM (middle) and RAM (bottom) for super-resolution with uncorrelated (left) and correlated (right) sources in 100 Monte Carlo runs. Sources 1 and 3 are coherent in the case of correlated sources. The area around the first two sources are zoomed in in each subfigure. Only results of the first 20 runs are presented for MUSIC for the purpose of better illustration.} \label{Fig:noisy_noiselevel}
\end{figure}

Our simulation results of 100 Monte Carlo runs are presented in Fig. \ref{Fig:noisy_noiselevel} (only the first 20 runs are presented for MUSIC for better illustration). In the absence of source correlations, MUSIC has satisfactory performance in most scenarios. However, its power spectrum exhibits only a single peak around the first two sources (i.e., the two sources cannot be separated) in at least 3 out of the first 20 runs (indicated by the arrows). Moreover, MUSIC is sensitive to source correlations and cannot detect source 1 when it is coherent with source 3. ANM cannot separate the first two sources in the uncorrelated source case and always produces many spurious sources. In contrast, the proposed RAM always correctly detects 4 sources near the true locations, demonstrating its capabilities in enhancing sparsity and high resolution. ANM and RAM take $0.87$s and $7.31$s on average, respectively, while these numbers can be greatly decreased with more sophisticated algorithms (see \cite{yang2014enhancing}).

\section{Conclusion} \label{sec:conclusion}
In this paper, we studied the spectral super-resolution problem with partial samples and MMVs. Motivated by its connection to the topic of LRMR, we proposed reweighted atomic-norm minimization (RAM) for achieving high resolution compared to currently prominent atomic norm minimization (ANM) and validated its performance via numerical simulations.

\bibliographystyle{IEEEtran}
%\bibliography{C:/Users/An&Xuan/Dropbox/myreferences1}
%\bibliography{J:/myreferences1}


\end{document}